\documentstyle[twoside,fleqn,espcrc2,epsf]{article}

\newcommand{\Tr}{\mbox{Tr}}
\newcommand{\be}{\begin{equation}}
\newcommand{\ee}{\end{equation}}
\newcommand{\bdm}{\begin{displaymath}}
\newcommand{\edm}{\end{displaymath}}

\newcommand{\StruveL}{{\mbox{\bf L}}}

\def\su3{$SU(3)$}

\def\spose#1{\hbox to 0pt{#1\hss}}
\def\ltapprox{\mathrel{\spose{\lower 3pt\hbox{$\mathchar"218$}}
 \raise 2.0pt\hbox{$\mathchar"13C$}}}
\def\gtapprox{\mathrel{\spose{\lower 3pt\hbox{$\mathchar"218$}}
 \raise 2.0pt\hbox{$\mathchar"13E$}}}
\def\inapprox{\mathrel{\spose{\lower 3pt\hbox{$\mathchar"218$}}
 \raise 2.0pt\hbox{$\mathchar"232$}}}

\title{
Finite-Volume Scaling of the Quenched Chiral Condensate\thanks{Presented by 
P.~H.~Damgaard}
}

\author{
Poul H.~Damgaard\address{
The Niels Bohr Institute, Blegdamsvej 17, DK-2100 Copenhagen, Denmark},
Robert G.~Edwards\address{
SCRI, The Florida State University, 
Tallahassee, FL 32306-4130, USA},
Urs M.~Heller$^{\rm b}$,
and Rajamani Narayanan\address{
American Physical Society,
One Research Road,
Ridge, NY 11961, USA}
}

\begin{document}

\begin{abstract}
In the large-volume limit $V\to\infty$ with $V \ll 1/m_{\pi}^4$
the mass-dependent chiral condensate is predicted to satisfy
exact finite-volume scaling laws that fall into three major 
universality classes. We test these analytical predictions with
staggered fermions and overlap fermions in gauge field sectors
of fixed topological charge $\nu$.   
\end{abstract}

\maketitle


\section{Finite-volume partion functions and RMT}

Surprisingly, there exists a deep connection between 
finite-volume gauge theories with spontaneous breaking of chiral
symmetries, and Random Matrix Theory (RMT). In the chiral limit 
$m_i\to 0$ such that
$\mu_i \equiv m_i\Sigma V$ are kept fixed, the field-theoretic
partition functions $Z_{\nu}(\{\mu_i\})$ become {\em identical} to
certain (chiral) RMT partition functions \cite{SV}. Here $\Sigma$ denotes
the (conventional) infinite-volume chiral condensate, and $\nu$ is 
the (fixed) topological charge of the gauge fields. This remarkable
identity of partition functions is stable under huge perturbations of
the RMT ``potential'', i.e. {\em universal} \cite{ADMN}.
RMT plays a role here because the
three major (chiral) classes of RMT ensembles correspond to 
the cosets of spontaneous chiral symmetry breaking. It is crucial
that one considers the limit $V\to\infty$ with $V\ll 1/m_{\pi}^4$.
Taken conversely, from reading off the scaling behavior of the 
finite-volume chiral condensate one deduces the coset of 
spontaneous chiral symmetry breaking! 

{}From the universal relationship between these different
expressions for the finite-volume partition functions follows that there
are also universal scaling relations for the chiral condensate and
higher chiral susceptibilities. Because the pertinent chiral RMT ensembles
are presumed relevant also for systems in condensed
matter physics, these universal scaling laws indeed go much beyond their
immediate context of four-dimensional Yang-Mills theories. 

The quenched limit can be taken in both the RMT framework
\cite{V} and that of the effective lagrangian \cite{OTV}. The
RMT framework provides a neat classification of the 
universality classes, and we shall employ this RMT language below.
We work in the quenched limit.

\section{Staggered Fermions}

At moderate gauge couplings
staggered fermions are almost blind to topology \cite{DHNR} (but see
also ref. \cite{Graz}), and we thus 
should compare analytical predictions to those of $\nu=0$ only. We have
performed a systematic analysis of all three different universality classes
in ref. \cite{DEHN}. The first study of this kind was performed by
Verbaarschot \cite{V} for the chUE universality class. 

\begin{figure}
\epsfxsize=2.8in
\centerline{\epsffile{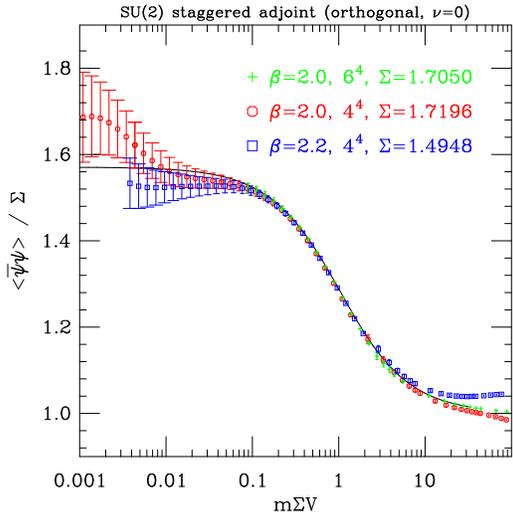}}
\vspace*{-8mm}
\caption{The rescaled quenched consensate for staggered fermions in the
the adjoint representation of SU(2).}
\label{fig:st_pbp_OE}
\end{figure}

There are exact analytical predictions for all three universality
classes \cite{DEHN}. We here concentrate on just
one of these, the chOE. 
For staggered fermions away from the continuum limit it corresponds to
all gauge groups SU($N_c\geq 2$) and fermions in the adjoint 
representation. The quenched chiral condensate for that case has the
analytical form (for $\nu=0$):
\begin{eqnarray}
 \frac{\Sigma_0^{\rm{\mbox chOE}}(\mu)}{\Sigma} =  
\mu I_{1}(\mu)K_{1}(\mu)
+ \frac{\pi}{2}\left[I_{0}(\mu)- \StruveL_{0}(\mu)\right] \cr
\;\; + \frac{\pi}{2}K_0(\mu)\left(\StruveL_0(\mu)I_1(\mu)-
\StruveL_1(\mu)I_0(\mu)\right] .
\label{sigmaOE2}
\end{eqnarray}
Eq. (\ref{sigmaOE2}) gives a parameter-free 
prediction for the scaling of $\Sigma_0(\mu)$ once $\Sigma$ is known.
We compare this prediction with lattice data in Fig.~\ref{fig:st_pbp_OE}.
Note the 
unusual behavior: The condensate {\em rises}
with decreasing mass, eventually saturating at the constant value
$\Sigma_0(\mu)/\Sigma = \pi/2$. Here, as $V$ is taken to infinity, 
the mass $m$ is taken to zero at
a rate proportional to $1/V$. This would normally imply that $\Sigma(\mu)
\to 0$ as $\mu\to 0$. The fact that it doesn't here
is due to the quenched approximation, and the $\nu=0$ sector. 

We have compared the analytical predictions for all three universality
classes using staggered fermions, and found remarkably good agreement,
even in rather small physical volumes \cite{DEHN}. One sees very clearly 
the striking scaling of the condensate: instead of depending on the
three parameters $m, \Sigma$ and $V$, it only depends on the scaling
combination $\mu=m\Sigma V$.

\section{Topology: Overlap Fermions}

It is far more interesting to test the analytical predictions
with lattice fermions that are sensitive to lattice gauge field topology.
Results have previously focused on the smallest eigenvalue distributions
\cite{EHKN}. We shall here present some of our results for the chiral
condensate, obtained using overlap fermions \cite{Neuberg98}. Similar
results for the case of the chUE universality class have been presented
in ref. \cite{HJL}. In contrast to staggered fermions, overlap fermions
have the additional advantage of belonging, at finite lattice spacing,
to precisely the same universality classes as continuum fermions.  

\begin{figure*}[t]
\vspace*{-5mm} \hspace*{-0cm}
\begin{center}
\epsfxsize = 0.8\textwidth
\centerline{\epsfbox[100 280 550 410]{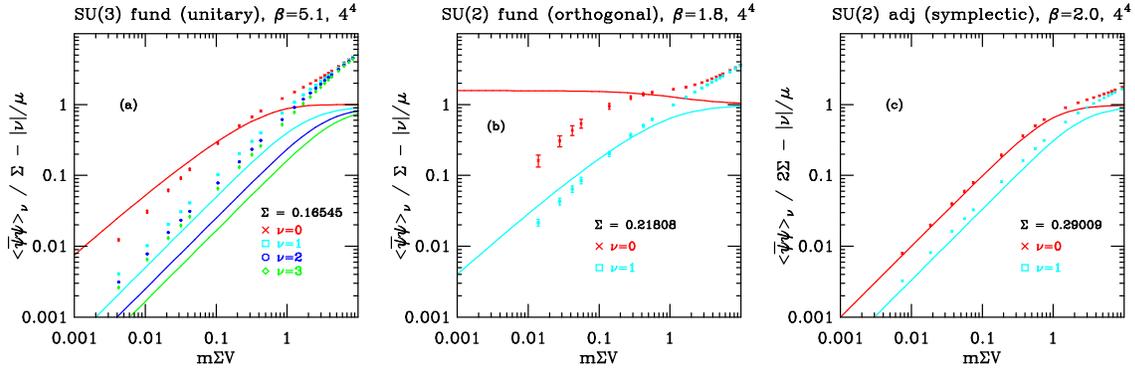}}
\end{center}
\caption{The rescaled quenched condensate
as a function of $\mu = m \Sigma V$ for overlap fermions in
(a) the fundamental representation of SU(3),
(b) the fundamental representation of SU(2),
(c) in the adjoint representation of SU(2).}
\label{fig:ov_pbp}
\end{figure*}

The massive overlap Dirac opertor is given by
\be
D(m)=\frac{1}{2} \left[1+m + (1-m) \gamma_5\epsilon(H_w(m))  \right] .
\ee
and the propagator for external fermions needs a subtraction to ensure
the proper chiral behavior in the massless limit,
\be
{\tilde D}^{-1}(m) = (1-m)^{-1} \left[ D^{-1}(m) -1 \right] .
\ee
The condensate $\Sigma(m) = \langle \Tr {\tilde D}^{-1}(m) \rangle$ can
then be easily obtained using Gaussian random source methods~\cite{ehn_pbp}.

Overlap fermions have exact zero modes in topologically non-trivial gauge
fields~\cite{EHKN}, and allow therefore comparison with RMT predicitions
in sectors with $\nu \ne 0$. We separate out the rather trivial
contribution from the exact zero modes and consider $\Sigma_\nu(\mu)/\Sigma
- |\nu|/\mu$. For its stochastic estimate, we use the fact that
$D^\dagger(m) D(m)$ commutes with $\gamma_5$, from which follows that the
contribution from the non-zero modes to $\Sigma_\nu$ is equal in both
chirality sectors so that we can choose to do the necessary inversion only
in the sector without zero mode~\cite{ehn_pbp}. Finally, using
\be
D^\dagger(m) D(m) = (1-m^2) \left[ D^\dagger(0) D(0) + \frac{m^2}{1-m^2}
 \right]
\ee
we can simultaneously compute for several masses using multiple Krylov
space solvers.

We considered examples for all three ensembles of the RMT
classification~\cite{DEHN}.
Our results are shown in Fig.~\ref{fig:ov_pbp}. In all cases the
condensate $\Sigma$ had been determined previously from the distribution
of the lowest non-zero eigenvalue~\cite{EHKN} and the RMT predicitions,
shown as curves in Fig.~\ref{fig:ov_pbp}, are therefore parameter free.

Due to the costs of the simulations with overlap fermions, we were restricted
to rather small volumes and relatively low statistics. As discussed in
detail in \cite{DEHN}, the deviations seen in Fig.~\ref{fig:ov_pbp} from the 
analytical predicitions are readily understood if one considers the
microscopic spectrum of the Dirac operator. One is effectively caught 
between the difficulty of probing the very smallest eigenvalue to very
high accuracy, and the finite-volume limitation that distorts the 
microscopic Dirac operator spectrum after just a few (averaged) eigenvalue
peaks. The best agreement with analytical predictions are found in the
case of the chSE universality class, in both the sector of $\nu=0$ and $\nu=1$.
This is again a reflection of the microscopic Dirac operator spectrum for
those gauge field configurations \cite{DEHN}. Even for the other two
universality classes the results are qualitatively in good agreement with
theory, in particular the shift due to higher topological charge sectors
is correctly reproduced.   

The work of R.G.E. and U.M.H. was partially supported by DOE contracts
DE-FG05-85ER250000 and DE-FG05-96ER40979, and the work of P.H.D.
by EU TMR grant ERBFMRXCT97-0122. Both P.H.D.
and U.M.H. also cknowledge support by NATO Science Collaborative Research
Grant CRG 971487. All the authors acknowledge the hospitality of the
Aspen Center for Physics, were this research was completed.

\vskip -3mm


\begin{thebibliography}{99}
%
\bibitem{SV}E.V. Shuryak and J.J.M. Verbaarschot, {\em Nucl. Phys. \/} 
{\bf A560} (1993) 306; M.A. Halasz and J.J.M. Verbaarschot, {\em Phys. Rev.
\/} {\bf D52} (1995) 2563; J.J.M. Verbaarschot, {\em Phys. Rev. Lett. \/} 
{\bf 72} (1994) 2531.
\bibitem{ADMN}G. Akemann, P.H. Damgaard, U. Magnea and S. Nishigaki,
{\em Nucl. Phys. \/} {\bf B487} (1997) 721.
\bibitem{V}J.J.M. Verbaarschot, {\em Phys. Lett. \/} {\bf B368} (1996) 137.
\bibitem{OTV}J. Osborn, D. Toublan and J.J.M. Verbaarschot, {\em Nucl. Phys.
\/} {\bf B540} (1998) 317; P.H. Damgaard, J. Osborn, D. Toublan and J.J.M.
Verbaarschot, {\em Nucl. Phys. \/} {\bf B547} (1999) 305.  
\bibitem{DHNR}P.H. Damgaard, U.M. Heller, R. Niclasen and K. Rummukainen,
hep-lat/9907019.
\bibitem{Graz} F. Farchioni, I. Hip and C.B. Lang, hep-lat/9907011.
\bibitem{DEHN}P.H. Damgaard, R.G. Edwards, U.M. Heller and R. Narayanan,
hep-lat/9907016.
\bibitem{EHKN}R.G. Edwards, U.M. Heller, J. Kiskis and R. Narayanan,
{\em Phys. Rev. Lett. \/} {\bf 82} (1999) 4188; F. Farchioni, I. Hip, 
C.B. Lang and M. Wohlgenannt, {\em Nucl. Phys.\/} {\bf B549} (1999) 364.
\bibitem{Neuberg98} H. Neuberger, {\em Phys. Lett. \/} {\bf B417} (1998) 141.
\bibitem{HJL}P. Hernandez, K. Jansen and L. Lellouch, hep-lat/9907022.

\bibitem{ehn_pbp} R.G. Edwards, U.M. Heller and R. Na\-ray\-anan,
{\em Phys. Rev. \/} {\bf D59} (1999) 094510.  

\end{thebibliography}
\end{document}